\documentclass[prb,aps,twocolumn,amsmath,amssymb,floatfix,superscriptaddress]{revtex4}

\usepackage[dvips]{graphics}
\usepackage{color}
\usepackage{soul}
\usepackage{mathrsfs}
\usepackage[colorlinks=true, citecolor=blue, urlcolor=blue]{hyperref}

\begin{document}

\title{Anomalous persistent current in a 1D dimerized ring with aperiodic site potential: Non-interacting and interacting cases}
	
\author{Souvik Roy}

\email{souvikroy138@gmail.com}
		
\affiliation{Physics and Applied Mathematics Unit, Indian Statistical Institute, 203 Barrackpore Trunk Road, Kolkata-700 108, India}
	
\author{Santanu K. Maiti}

\email{santanu.maiti@isical.ac.in}
	
\affiliation{Physics and Applied Mathematics Unit, Indian Statistical Institute, 203 Barrackpore Trunk Road, Kolkata-700 108, India}

\author{David Laroze}

\email{dlarozen@uta.cl}

\affiliation{Instituto de Alta Investigaci\'{o}n, Universidad de Tarapac\'{a}, Casilla 7D, Arica, Chile}
	
\begin{abstract}

In this work, we investigate the magnetic response by examining flux-driven circular currents in a Su-Schrieffer-Heeger (SSH) tight-binding
(TB) ring threaded by an Aharonov-Bohm (AB) flux, $\phi$. We consider both non-interacting and interacting electrons, where site energies
are modulated by a slowly varying cosine form. Repulsive electron-electron interaction is incorporated through an on-site Hubbard term, 
and we analyze the system using the Hartree-Fock (HF) mean-field (MF) approximation. We discuss the characteristics of flux-driven circular
currents to aperiodic potentials, dimerized hopping integrals, and Hubbard interactions. For the chosen aperiodic potential, both the 
strength and configuration play a crucial role, and we explore these aspects in depth. Interestingly, we observe a counterintuitive
delocalizing effect as the aperiodic potential increases, unlike in conventional disordered rings. The effects of system size, filling 
factor, the presence of circular spin current, and the accuracy of MF results are also discussed. Finally, we provide a brief description 
of possible experimental realizations of our chosen quantum system. This investigation can be extended to explore additional properties 
in various loop substructures, promising further insights.

\end{abstract}

\maketitle

\section{Introduction}

The study of localization phenomena in various types of disordered systems was pioneered by P. W. Anderson's seminal work in 1958~\cite{r1}.
Anderson's findings showed that a one-dimensional (1D) system becomes fully localized when exposed to a random (uncorrelated) disorder with
the critical disorder strength being effectively zero~\cite{r2,r3}. This result rendered such systems rather trivial in nature~\cite{r4}.
Later, an interesting model referred to as the Aubry-André-Harper (AAH) model realized that the disorder is deterministic,
following a specific pattern rather than being random. In this model, the critical disorder strength is finite~\cite{r5,r6,r7}, leading to
distinct phases: the system is conductive or insulating depending on whether the disorder strength is below or above this
threshold~\cite{r8,r9,r10,r11,r12,r13,r14}. The AAH model has sparked significant interest within the research community for its unique
and valuable insights into electronic and phononic transport across various physical systems~\cite{r15,r16,r17,r18,r19,r20,r21,r22}.

Various types of AAH models are widely studied to explore physical phenomena in condensed matter systems and beyond. The most common form
for site energy in tight-binding notation is: $\epsilon_{p} = W \cos(2\pi b p + \phi_{\lambda})$, where $W$ denotes the strength, $p$ is
the site index, $b$ is an irrational number, and $\phi_{\lambda}$ represents the AAH phase. The irrationality of $b$ induces correlated
disorder, which can be selectively tuned by adjusting the phase factor $\phi_{\lambda}$. When $W = 0$, the system is ideal (disorder-free).
In this work, we consider a distinct form of cosine modulation where the site energies are given by $\epsilon_{p} = W \cos(2\pi p^{\nu})$.
The parameter $\nu$ appears as the exponent of the site index $p$, and it can vary between $0$ and $1$, while the factor $b$ is absent.
When $\nu$ is either $0$ or $1$, all site energies are identical, and the system becomes fully ordered. For values of $\nu$ within
$0 < \nu < 1$, however, the system becomes aperiodic, commonly referred to as a slowly varying disorder~\cite{r23,r24,r25,r26,r27,r28}.
Systems with this type of aperiodic distribution have been compelling subjects of recent research, engaging scientists across various
fields. The complexity and intrigue further increase when hopping integrals are constrained according to the Su-Schrieffer-Heeger (SSH)
model~\cite{r29,r30,r31}, where hopping strengths exhibit dimerization.
\begin{figure}[ht]
{\centering\resizebox*{6.5cm}{3cm}{\includegraphics{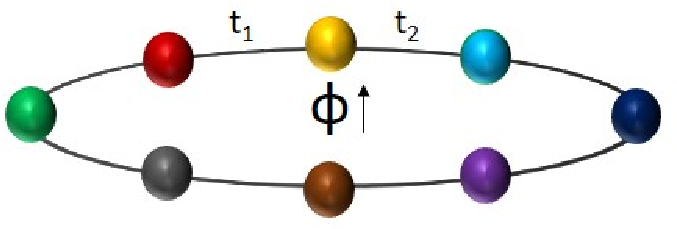}\par}}
\caption{(Color online). Schematic diagram of a quantum ring threaded by an AB flux $\phi$. The hopping strengths are 
dimerized, and the ring is subject to an aperiodic potential. Different colored balls signify that site energies vary 
from one site to the other. In the presence of flux $\phi$, a net circular current is established in the ring.}
\label{f1}
\end{figure}
A few studies have investigated the re-entrant delocalization phenomenon in quasi-periodic systems based on the SSH model~\cite{r32,r33,r34}. 
However, to our knowledge, studies on the combined effect of slowly varying site energy with SSH-type hopping dimerization remain relatively limited. In this work, we investigate the magnetic response of a mesoscopic ring under an Aharonov-Bohm flux, $\phi$, to unravel the interplay between this specific type of AAH potential and hopping dimerization on electron transport. It was proposed quite a long time back that a ring with a circumference less than or comparable to the electronic mean free path can carry a net circular current when threaded with a magnetic flux~\cite{r35,r36,r37}. This is called a flux-driven circular current, and it persists even when the flux is removed. Later, numerous experiments and theoretical studies~\cite{r38,r39,r40,r41,r42,r43,r44,r45,r46,r47,r48,r49,r50,r51,r52,r53,r54,r55,r56,r57,r58} have been conducted to explore various characteristics of flux-driven circular currents under diverse physical conditions. Nonetheless, {\em the model we are dealing with is relatively new, and the study of magnetic responses through the computation of persistent current in the presence of such slowly varying disorder and SSH-type hopping dimerization is uncommon in the literature}. As the circular current is directly linked with the nature of eigenstates, the combined effect of slowly varying site potential and hopping dimerization on conducting properties can be understood quite clearly. We aim to investigate various engrossing phenomena both for the non-interacting and interacting electrons. Here, it is relevant to note that truly capturing the effects of a `slowly varying' disorder distribution requires considering a significantly large system size. However, in such cases, finding flux-driven persistent current becomes challenging, as it is a mesoscopic phenomenon. Therefore, we must compromise on the ring size. We believe, though, that even smaller rings can effectively capture all relevant aspects of the slowly varying aperiodic potential. 

Our chosen quantum ring is described within a tight-binding framework, and it comprises both the aperiodic site energies and the hopping dimerization. The repulsive electron-electron interaction is included in the Hamiltonian via the on-site Hubbard term~\cite{r59,r60,r61,r62,r63,r64,skm1}. Working with an interacting system is always a difficult task, especially when considering calculations using the full many-body (MB) Hamiltonian. The dimension of the many-body (MB) Hamiltonian matrix increases sharply--almost exponentially--with system size and the number of up and down spin electrons. In such cases, we must either restrict ourselves to smaller system sizes or adopt alternative approaches. Very small systems sometimes fail to capture the full characteristics, especially the signature of the chosen aperiodic potential, necessitating methods beyond calculations with the full MB Hamiltonian. One effective technique is the Hartree-Fock (HF) mean-field (MF) approach~\cite{r59,r60,r61,r62}, where the MB Hamiltonian is approximated as a sum of effective up-spin and down-spin Hamiltonians. From there, the remaining calculations are similar to the non-interacting case, allowing us to handle significantly larger system sizes. Here, it is relevant to note that MF calculations may overlook certain crucial phenomena. While the Bethe ansatz~\cite{r62,ba1} offers an alternative solution for the 1D Hubbard model, its applicability is limited in the presence of disorder. Since the present work focuses on the combined effects of disorder and electron-electron interactions, relatively larger rings are required.
However, due to the inapplicability of the standard Bethe ansatz in disordered systems, we adopt the HF-MF technique instead.  

From our extensive numerical analysis, we find that irrespective of e-e interaction, the interplay between dimerized hopping and aperiodic potential is highly interesting. Similar to the conventional AAH model, where the change of phase factor $\phi_{\lambda}$ leads to a separate disordered configuration, in our chosen system, the parameter $\nu$ does. But, there is a crucial difference. Setting $W$ at a particular value, if we tune $\nu$, the effective disorder strength gets modified, and hence, a change in transport behavior is naturally expected. Unlike the conventional disordered case, we find that under certain physical conditions, the disorder can enhance the current. This re-entrant delocalizing behavior is further confirmed through the analysis of the {\em normalized participation ratio} (NPR)~\cite{r19,npr1,npr2}, where eigenstates, rather than eigenenergies, are considered. Our system also exhibits another intriguing phenomenon -- the presence of almost flat energy levels to the magnetic flux $\phi$ near the energy band edges. This feature can be exploited to selectively control the current depending on electron filling. Additionally, we observe a nonzero circular `spin current' under specific conditions. All these issues might be very interesting in transport phenomena, and have not been explored earlier.    

The rest of the paper is outlined as follows. In Sec. II, first, we describe the model quantum ring system within a TB framework, in the presence of slowly varying site potentials and two different kinds of hopping integrals in the form of an SSH model. Then, we illustrate the HF-MF scheme to work with the Hubbard interaction. In Sec. III, we present and discuss all the essential results for both non-interacting and interacting ring systems, including detailed band structures, circular charge and spin currents, and an NPR study. Finally, in Sec. IV, we conclude our findings.  
       
\section{Quantum ring and theoretical description}

This section describes the quantum ring within the tight-binding (TB) framework, followed by a self-consistent prescription for decoupling the interacting Hamiltonian. The theoretical steps to calculate the ground state energy and persistent current are presented in the latter part of this section.

\subsubsection{TB Hamiltonian of the quantum ring}

The quantum ring with $N$ atomic sites, threaded by an AB flux $\phi$, is schematically shown in Fig.~\ref{f1}. The colored balls represent the atomic sites in the ring, and different colors are used to indicate that the site potentials are different. Within the nearest-neighbor hopping (NNH) approximation of the TB Hamiltonian of the ring is written as
\begin{align}
	H&= \sum_{p,\sigma}\epsilon_{\sigma,p}c_{\sigma,p}^\dagger c_{\sigma,p}
	+\sum_{p(odd),\sigma}(t_{1}e^{i\theta}c_{\sigma,p}^\dagger c_{\sigma,p+1}+h.c.)\nonumber\\& +\sum_{p(even),\sigma} (t_{2}e^{i\theta}c_{\sigma,p}^\dagger c_{\sigma,p+1}+h.c.)
	+\sum_{p,\sigma}U n_{\uparrow,p} n_{\downarrow,p}.
	\label{Eq 1}
\end{align}
The above Hamiltonian mainly comprises three parts: site energy, electron hopping, and Coulomb interaction. In the first term, the slowly varying site energy is incorporated with $\epsilon_{\sigma,p}$ ($\sigma=\uparrow \textnormal{or}\downarrow$ and $p$ is the lattice site index). The site energy $\epsilon_{\sigma,p}$ for both the up and down spin electrons are taken as $W\cos(2\pi p^\nu)$, where $W$ is the strength 
of the disorder, and $\nu$ is the parameter of slowly varying site potential~\cite{r26,r27,r28}. When $\nu=0$ (or $1$), all the site energies are identical to each other, and the ring becomes impurity-free, which is referred to as a perfect ring. However, for intermediate values of $\nu$ ($0<\nu<1$), the site energies are different following the above-mentioned expression, and the ring becomes a disordered (correlated) one. The second and third terms are associated with electron hopping, where to consider hopping dimerization, we incorporate two different hopping strengths on alternate bonds that are specified by the parameters $t_1$ and $t_2$, respectively. Due to the magnetic flux $\phi$, a phase factor $\theta$ appears in the hopping terms, and it is expressed as $\theta=2\pi\phi/N$ ($N$ denotes the total number of lattice sites in the ring). The magnetic flux is measured in units of the elementary flux-quantum $\phi_0$ ($=ch/e$, where $c$, $h$, and $e$ are the fundamental constants). In the last term, we consider Coulomb (electron-electron) interaction~\cite{r55,r56,r57,r58}, with its strength specified by $U$, and $n_{\uparrow,p}$ (or $n_{\downarrow,p}$) represents the occupation index for up (or down) spin electrons. 

\subsection{The self-consistent mechanism under the mean field scheme}

Mean-field (MF) theory~\cite{r59,r60,r61,r62} offers a convenient way for decoupling the interacting tight-binding Hamiltonian, and 
this theory is very helpful, especially for large system sizes compared to others, as already mentioned earlier. 
Under MF scheme, the MB term $Un_{\uparrow,p} n_{\downarrow,p}$ can be expressed in single electron picture as   
\begin{align}	
Un_{\uparrow,p} n_{\downarrow,p}=U(\langle n_{\uparrow,p}\rangle n_{\downarrow,p}+\langle n_{\downarrow,p}\rangle n_{\uparrow,p}-\langle n_{\uparrow,p}\rangle \langle n_{\downarrow,p}\rangle).
\end{align}
Using this form, the TB Hamiltonian becomes
\begin{align}
	H&= H_{\uparrow}+H_{\downarrow}-\sum_{p}U\langle n_{\uparrow,p}\rangle \langle n_{\downarrow,p}\rangle.
\end{align}
where $H_{\uparrow}$ and $H_{\downarrow}$ are the sub-Hamiltonians associated with up and down spin electrons. These sub-Hamiltonians
are explicitly expressed as 
\begin{subequations}
	\begin{align}
	H_{\uparrow}= \sum_{p,\uparrow}(\epsilon_{\uparrow,p}+U\langle n_{\downarrow,p}\rangle)c_{\uparrow,p}^\dagger c_{\uparrow,p}
	+\sum_{p,odd,\uparrow}(t_{1}e^{i\theta}c_{\uparrow,p}^\dagger c_{\uparrow,p+1}+\nonumber\\h.c.) +\sum_{p,even,\uparrow} (t_{2}e^{i\theta}c_{\uparrow,p}^\dagger c_{\uparrow,p+1}+h.c.),
		\\
	H_{\downarrow}= \sum_{p,\downarrow}(\epsilon_{\downarrow,p}+U\langle n_{\uparrow,p}\rangle)c_{\downarrow,p}^\dagger c_{\downarrow,p}
	+\sum_{p,odd,\downarrow}(t_{1}e^{i\theta}c_{\downarrow,p}^\dagger c_{\downarrow,p+1}+\nonumber\\h.c.) +\sum_{p,even,\downarrow} (t_{2}e^{i\theta}c_{\downarrow,p}^\dagger c_{\downarrow,p+1}+h.c.).
		\label{Eq 3}
	\end{align}
\end{subequations}
We can start with initial estimates for $n_{\uparrow,p}$ and $n_{\downarrow,p}$ to obtain a self-consistent solution, leading to a convergent solution where we acquire new sets of these parameters. 

Getting the converged $H_{\uparrow}$ and $H_{\downarrow}$, we find the energy eigenvalues of the up and down spin electrons by diagonalizing them. Then, the ground state energy is obtained by summing over the appropriate energy levels associated with up and down spin electrons via the relation  
\begin{align}
E_{0}&=E_{\uparrow}+E_{\downarrow}-\sum_{p}U \left<n_{\uparrow,p}\right>\left<n_{\downarrow,p}\right>,
\label{Eq 4}
\end{align}
where the sum of the eigenenergies of up and down spin electrons up to Fermi level is denoted by $E_{\uparrow}$ and $E_{\downarrow}$, respectively. The first order derivative of $E_0$ provides the persistent current~\cite{r36} in the ring system, and the expression of it is
\begin{center}
\begin{align}
\hskip 0.2in I_{\phi}= -c\frac{\partial E_{0}(\phi)}{\partial \phi}.
\label{Eq 5}
\end{align}
\end{center}
The current expression suggests that the nature of $E_0(\phi)$ is extremely important, and this ground state energy, on the other hand, is directly involved in the available up and down spin energy levels.

From Eq.~\ref{Eq 5}, we determine the circular charge current. Similarly, we can calculate individual current components, namely, the currents associated with up and down spin electrons, and then obtain the circular spin current. Taking the first-order derivative of the ground state energy to magnetic flux for up-spin electrons, we get the up-spin current, denoted as $I_{\mbox{\small up}}$. Likewise, the contribution from down-spin electrons gives the down-spin current, referred to as $I_{\mbox{\small down}}$. The difference between these two components yields the circular spin current, i.e., $I_{\mbox{\small spin}}=I_{\mbox{\small up}}-I_{\mbox{\small down}}$. By summing these two components, we can also calculate the circular charge current, which exactly matches the one determined from Eq.~\ref{Eq 5}. Similar to the spin current, the charge current can be referred to as $I_{\mbox{\small charge}}$. However, among the two currents, since the charge current is mostly discussed, we denote it simply as `$I$' instead of writing $I_{\mbox{\small charge}}$. For the spin current, we explicitly use $I_{\mbox{\small spin}}$. 

{\em Estimation of localization from eigenstates}: For an $n$th normalized eigenstate, the normalized participation ratio is defined as~\cite{r19,npr1,npr2}
\begin{equation}
NPR_n=\left(N \sum_p |\psi_p^n|^4\right)^{-1}.
\label{eqnpr}
\end{equation}
For an extended state, NPR approaches unity, whereas, for a localized state, it becomes very small, nearly zero.

\section{Numerical results and discussion}

In this section, we have placed the essential results for non-interacting and interacting cases. For a non-interacting picture, the effects of disorder strength, slowly varying parameters, and hopping dimerization are extensively examined, while for an interacting picture, the Hubbard interaction parameter $U$ is taken into consideration along with the other parameters. We first discuss the non-interacting scenario before concentrating on the interacting one. The current is calculated in the unit of $\mu A$, while the variables like disorder strength, hopping integrals, and Coulomb interaction are measured in the unit of $eV$. We choose two values,
$1.2$ and $0.8$, to define the inequality between $t_1$ and $t_2$. Unless it is categorically mentioned, the current is referred to as the charge current. When the current is computed for a particular value of magnetic flux, we choose $\phi=0.2\phi_{0}$. In other cases, we vary the flux in a wide window. The behavior of spin current and the conducting nature of different eigenstates in terms of NPR are also briefly discussed in the appropriate parts. Below, we present all the results one by one in different sub-sections. 

\subsection{Non-interacting case}

\subsubsection{Energy band diagram}

For perfect and slowly changing disordered situations at two distinct conditions
\begin{figure}[ht]
\noindent
{\centering\resizebox*{8.5cm}{7.5cm}{\includegraphics{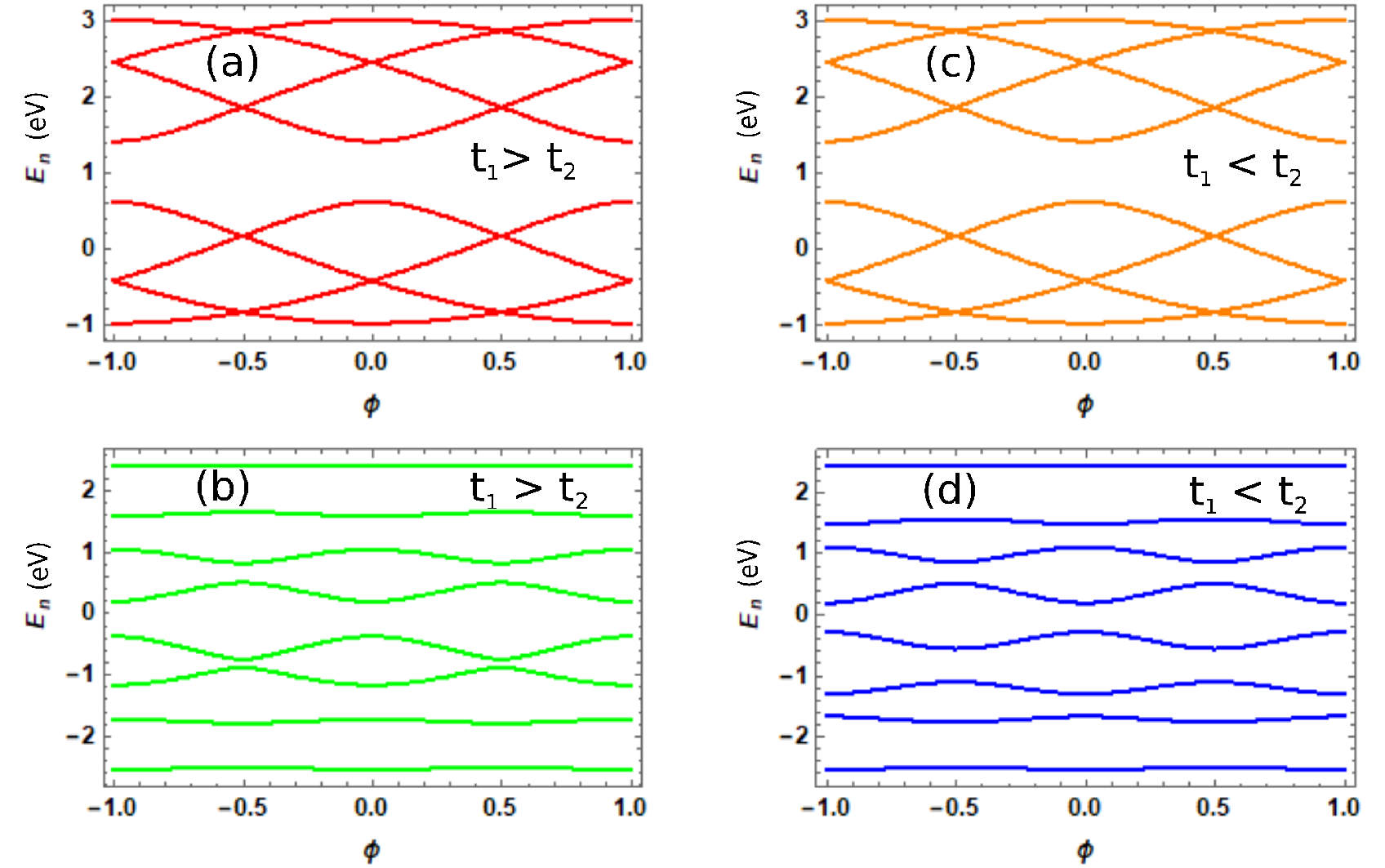}\par}}
\caption{(Color online). Variation of energy eigenvalues with AB flux $\phi$ for the interaction-free ring. The first and second rows correspond to $\nu=0$ and $0.3$, respectively. Two different columns are associated with two different dimerized conditions, viz, $t_1>t_2$ and $t_1<t_2$. Here, we choose $N=8$ and $W=1$.}
\label{f2}
\end{figure}
between the hopping integrals, we plot the energy band diagrams with magnetic flux in Fig.~\ref{f2}. We get two sub-bands and almost mutually canceling slopes for $t_1>t_2$ and $\nu=0$, therefore, the current will be minimal when estimated at half-filling. Energy levels that fluctuate with flux and do not cancel out the slopes of neighboring ones might be expected to produce current, depending on the filling. 
However, in the plot (Fig.~\ref{f2} (a)), such slopes from different energy levels will be mutually canceled to 
each other, especially at half-filling. Similar kinds of characteristics also occur for $t_{1}<t_{2}$ at $\nu=0$.
\begin{figure}[ht]
\noindent
{\centering\resizebox*{6cm}{10cm}{\includegraphics{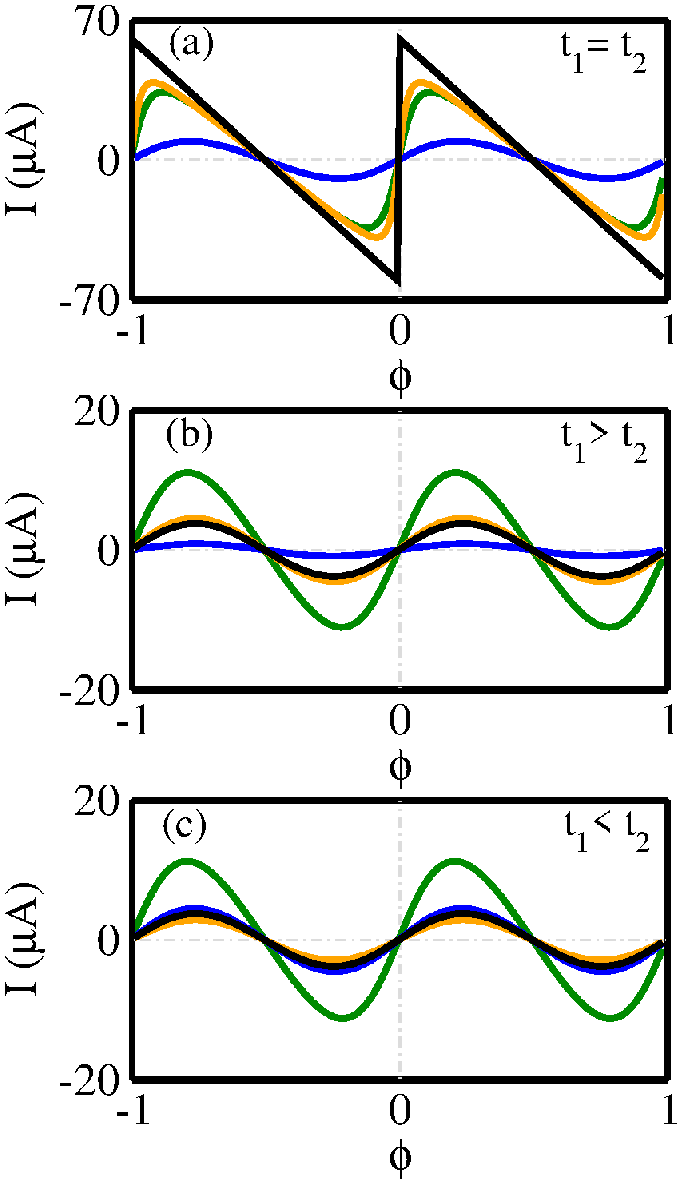}\par}}
\caption{(Color online). Dependence of current on magnetic flux $\phi$ at half-filling for the non-interacting case ($U=0$). Three different sub-plots from top to bottom correspond to $t_1=t_2$, $t_1>t_2$, and $t_1<t_2$, respectively. In each sub-figure, the green, blue, 
orange, and black curves are for $\nu=0.3$, $0.6$, $0.9$, and $0$ (or $1$), respectively. Here, we set $W=1$ and $N=16$.}
\label{f3}
\end{figure}
\begin{figure}[ht]
\noindent
{\centering\resizebox*{6cm}{10cm}{\includegraphics{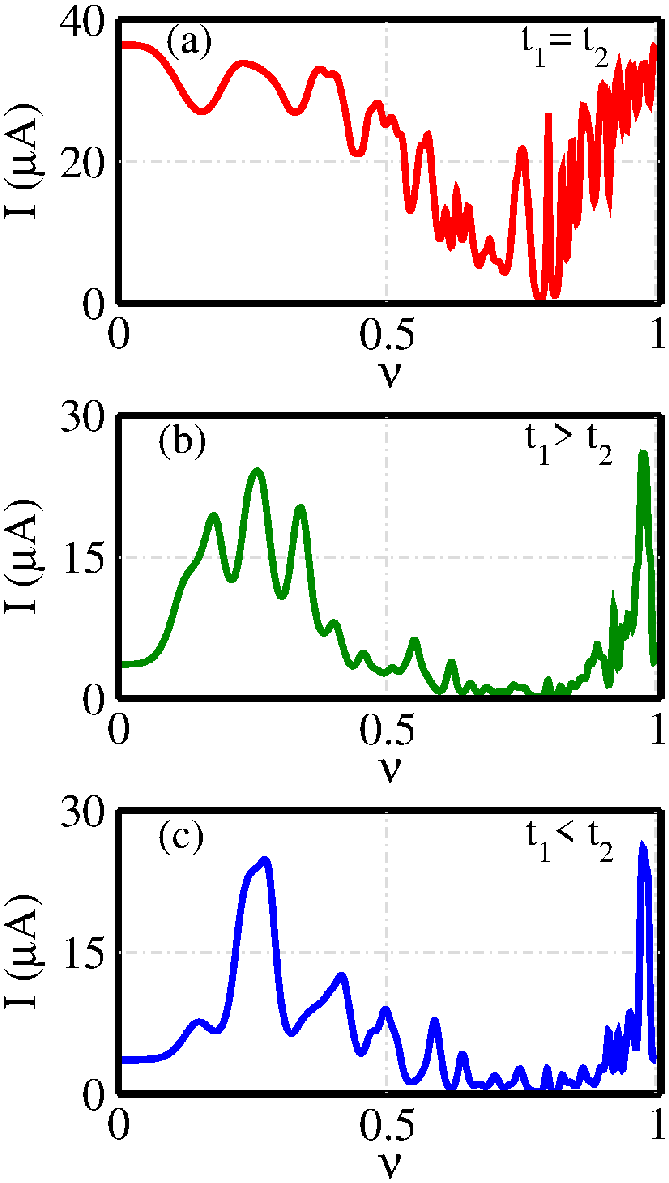}\par}}
\caption{(Color online). Dependence of current on the slowly varying modulation index $\nu$ for the non-interacting rings in the half-filled band case under three distinct conditions among $t_1$ and $t_2$. The ring size and the AAH modulation strength remain the same as used in Fig.~\ref{f3}.}
\label{f4}
\end{figure}
\begin{figure}[ht]
\noindent
{\centering\resizebox*{6cm}{10cm}{\includegraphics{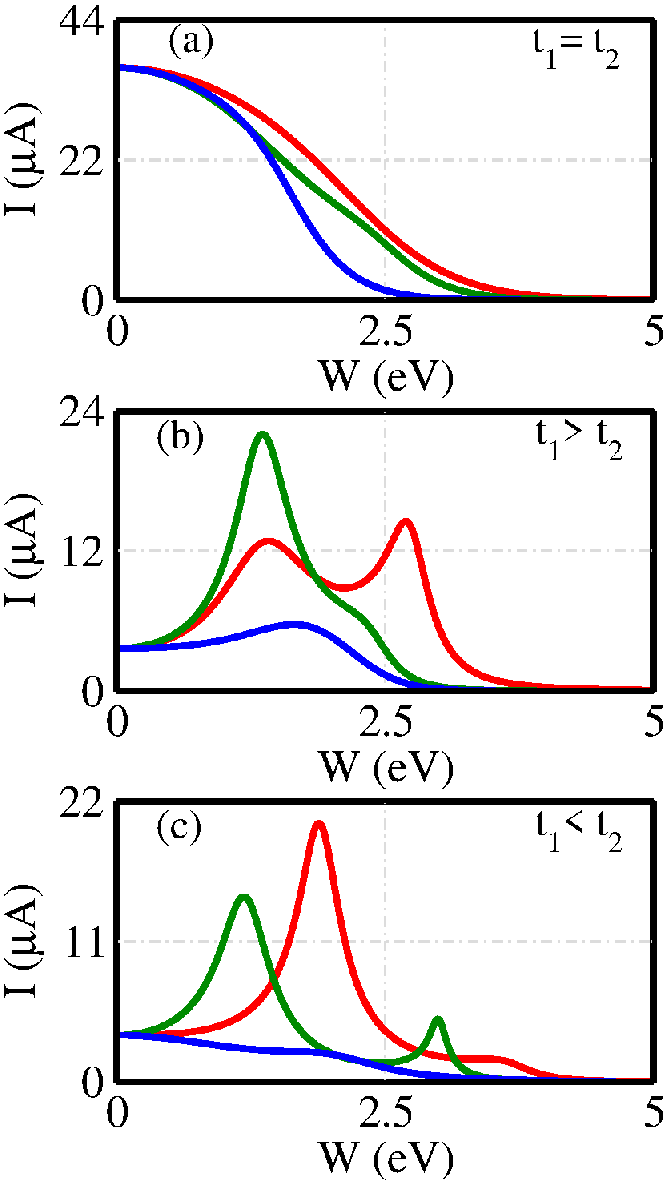}\par}}
\caption{(Color online). Variation of current with disorder strength $W$ for non-interacting rings in the limit of half-filling. In each sub-figure, the red, green, and blue curves are for $\nu=0.1$, $0.3$, and $0.9$, respectively. The system size remains unchanged as taken in Fig.~\ref{f3}.}
\label{f5}
\end{figure}
The plots corresponding to the red and orange colors in Fig.~\ref{f2} clarify these two scenarios. Furthermore, when we examine the green and blue curves for $\nu=0.3$ in the figure, we notice that the energy levels at the band edges exhibit an almost flat character. The number of such nearly flat bands increases with system size. For any conventional ring (viz, perfect and random (uncorrelated) disordered rings), such a situation with nearly flat bands is often not present. Additionally, the distance between the sub-bands appears to be less when we focus near the band center, and the slopes of the energy levels do not precisely cancel with each other. Due to the interplay between the hopping dimerization and site energy disorder, these energy levels close to the band center are significantly impacted. As a result, one may anticipate that such energy levels will produce a sufficient quantity of current that is higher than the previous one. In light of this, the combined impact of hopping dimerization and slowly varying quasi-periodicity on the band spectrum motivates us to pursue and uncover additional intriguing properties of the current.
\begin{figure*}[ht]
\noindent
{\centering\resizebox*{12cm}{11cm}{\includegraphics{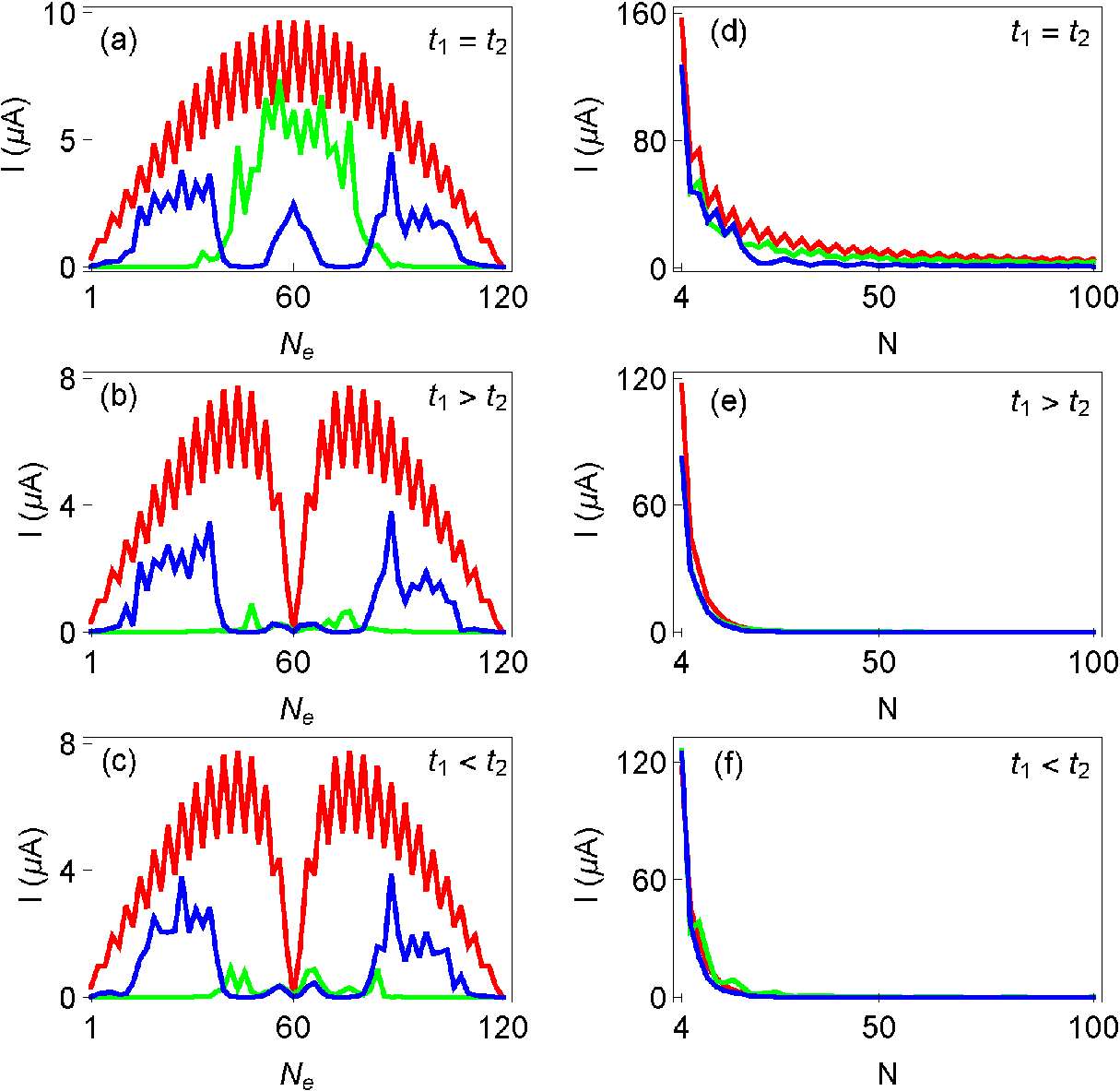}\par}}
\caption{(Color online). Left column: Circular current as a function of total number of electrons $N_e$ for a $60$-site ring. Right
column: Circular current as a function of ring size in the half-filled band case. In each sub-figure, the red, green, and blue curves
are for $\nu=0$, $0.5$, and $0.8$, respectively. The results are worked out for the non-interacting rings, considering $W=1$.}
\label{f6}
\end{figure*}
\begin{figure*}[ht]
\noindent 
{\centering\resizebox*{12cm}{13cm}{\includegraphics{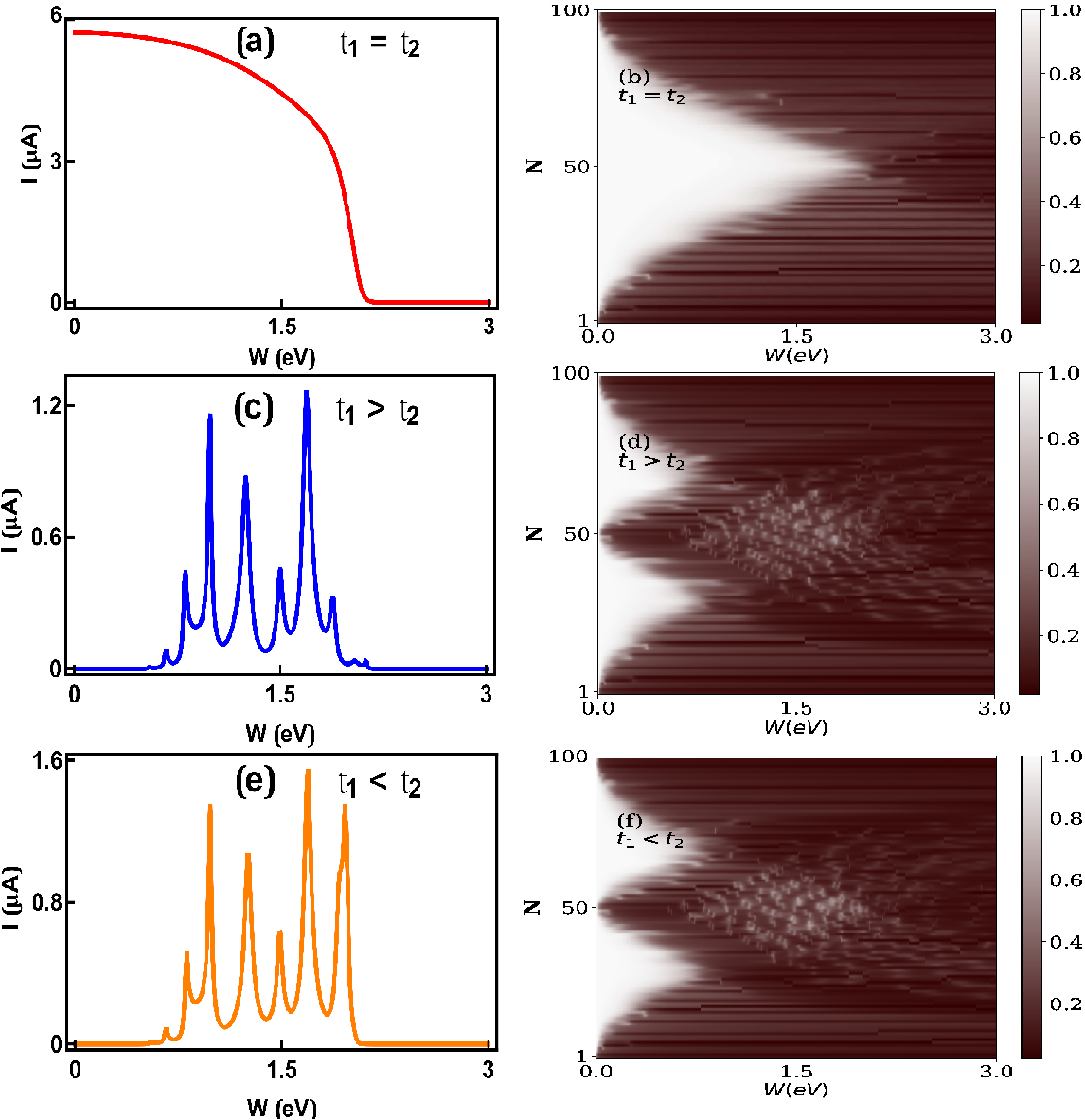}\par}}
	\caption{(Color online). First column: Variation of the typical current (at half filling) as a function of $W$. Second column: Density plot of NPR values as functions of eigenstate index and disorder strength $W$. Here, we choose a non-interacting ring with $N=100$, $\nu = 0.3$, and $\phi=0.2\phi_{0}$. Since $U = 0$, the eigenvalues and NPRs are identical for both up and down spin electrons. Therefore, we average the NPRs for up and down spin electrons to ensure that the NPR stays within its highest permissible limit of $1$. The reduction and/or enhancement of current with $W$ is interrelated with the nature of the contributing energy eigenstates, and it can be understood from the behavior of NPRs. The color bar denotes the NPR values, ranging from $0$ to $1$.}
\label{rev2}
\end{figure*}

\subsubsection{Variation of current with magnetic flux $\phi$}

In Fig.~\ref{f3}, the systematic variation of current with $\phi$ at constant disorder strength and filling is shown. The top, 
middle, and lower panels correspond to the cases with $t_{1}=t_{2}$, $t_{1}>t_{2}$, and $t_{1}<t_{2}$, respectively. In each plot, various colored curves are associated with different $\nu$ values. The current magnitude for $\nu=0$ (or $1$) is at its highest when $t_{1}=t_{2}$ compared to all other values of $\nu$. The curve containing black color in Fig.~\ref{f3}(a) provides a clear indication of 
such behavior. The system is disordered, and the current is smaller than in the preceding instance for the other values of $\nu$. However, the current is greater in certain disordered cases compared to the ideal one for the other two situations ($t_{1}>t_2$ and $t_{1}<t_{2}$). In such a non-trivial scenario, the presence of a slowly varying potential alone is not enough. Thus, the outcome of the first plot is easy to follow. The hopping dimerization, together with the slowly changing potential, will play a crucial part in this sort of behavior when we impose the circumstances between the hopping integrals. 

\subsubsection{Current with $\nu$}

The above results suggest that it would be meaningful to continually alter $\nu$ and plot the current with $\nu$ at some fixed $\phi$ 
and $W$, to derive further intriguing scenarios. These characteristics are discussed in this sub-section. In Fig.~\ref{f4} we present the 
results, and they are computed for three alternative conditions between the hopping integrals as earlier. The current for $\nu=0$ ($1$) is greater than that for other values of $\nu$, when $t_1=t_2$. The result is somewhat ordinary because $\nu=0$ and $1$ represent a perfect lattice, where the current is expected to be higher than in the disordered situations. If we set
$t_{1}\ne t_{2}$, the situation is different. In Figs.~\ref{f4}(b) and (c), the outcomes for $t_{1}>t_{2}$ and $t_{1}<t_{2}$ are shown. In certain situations, the current system favors predominantly disordered instances over those without disorders. The alteration of $\nu$ modifies the distribution of site energies within the system, thereby influencing the energy eigenvalues and consequently, the current. Additionally, the localizing (lesser current) and de-localizing (higher current) phenomena of the system are shown by the current, which oscillates (growing and reducing) with $\nu$ under all situations between the hopping integrals. These graphs enable us to observe how the current fluctuates between higher and lower levels, identifying several transitions between localization and de-localization under the change of $\nu$.

\subsubsection{Dependence of current with slowly varying modulation strength $W$}

We plot the current with $W$ in Fig.~\ref{f5} for three circumstances between the hopping integrals at the half-filled band case to 
thoroughly examine the relationship between disorder strength and current. In each figure, the 
curves with red, green, and blue colors show three distinct values of $\nu$, which are $0.1$, $0.3$, and $0.9$, 
respectively. Three separate rows reflect these three different circumstances involving the hopping integrals. For whatever value of $\nu$, the current always decreases with $W$ when $t_1=t_2$. As a result, for such a condition between the hopping integrals, the localization scenario happens with an increase in disorder strength, but there is no indication of delocalization or a re-entrant delocalizing scenario. The non-trivial signature, however, appears when $t_{1}\ne t_{2}$. While the current rises and lowers twice for $t_{1}>t_{2}$ at $\nu=0.1$, it displays two unique peaks. A similar kind of behavior occurs for the other values of $\nu$ as well, however, the second peak is more significant for $\nu=0.1$. For $t_{1}<t_{2}$, scenarios of a similar kind are shown. The secondary peak occurs for $\nu=0.1$ and $0.3$. The interplay between the slowly varying disorder and hopping dimerization leads to this increase in current and re-entrant delocalization. Therefore, in the presence of hopping dimerization in various ranges of disorder strength, we may change the conducting characteristics of the system by adjusting $W$ and selecting certain unique values of $\nu$. 
\begin{figure}[ht]
\noindent
{\centering\resizebox*{7.5cm}{11cm}{\includegraphics{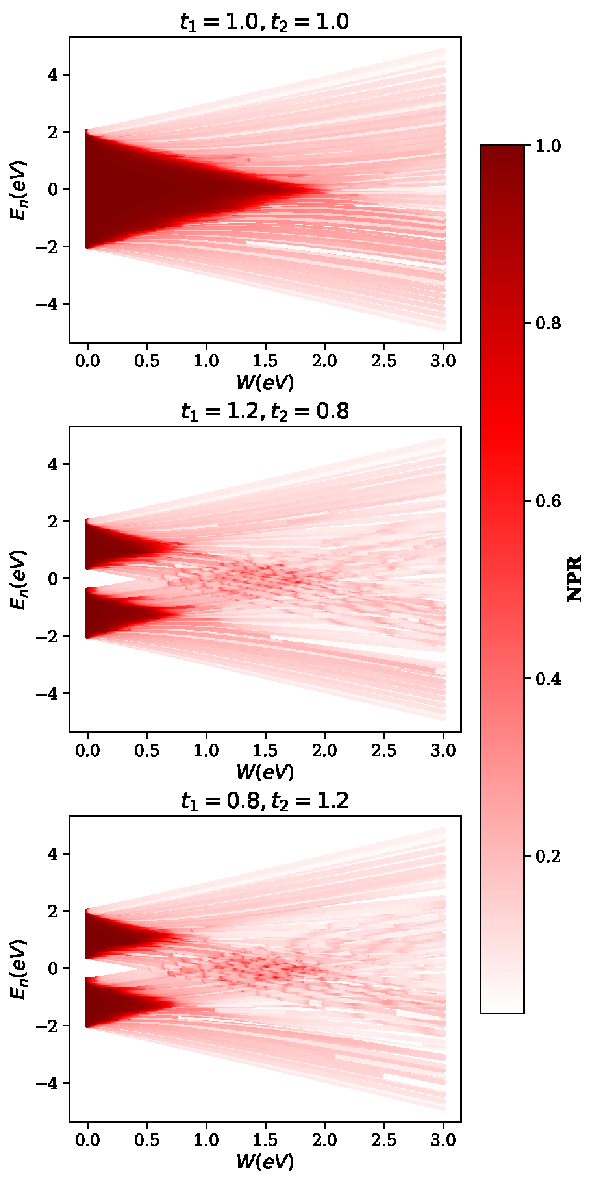}\par}}
\caption{(Color online). Density plot of NPRs as functions of energy eigenvalues and disorder strength $W$ for the identical ring as used in Fig.~\ref{rev2}. As in this case, $U = 0$, the eigenvalues and NPR remain the same for both up and down spin electrons. Consequently, we take the average NPR over both spin components, ensuring that its maximum possible value remains constrained to $1$. The color bar indicates the NPR values of different eigenstates, ranging from $0$ to $1$.}
\label{revband}
\end{figure}
\begin{figure}[ht]
\noindent
{\centering\resizebox*{8cm}{7cm}{\includegraphics{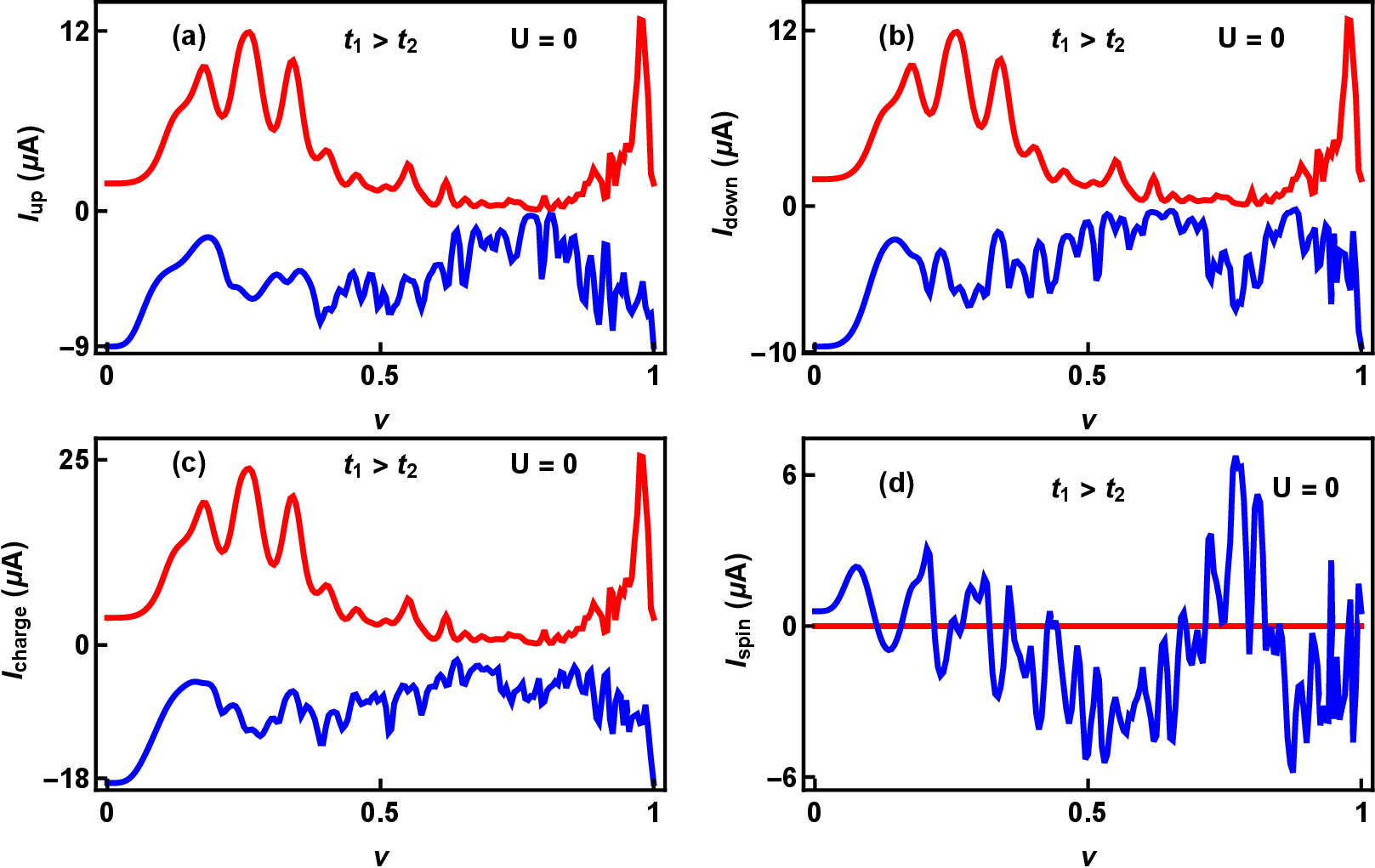}\par}}
\caption{(Color online). Variation of up and down spin currents, along with the associated charge and spin currents, as a function of $\nu$ for both half-filled (red) and non-half-filled (blue) cases, in the absence of electron-electron interaction. In the non-half-filled case, the system consists of $7$ up-spin and $5$ down-spin electrons. For the half-filled limit, an equal number of up and down spin electrons exist, like in the previous cases. Here, we choose a $16$-site ring, with the condition $t_1 > t_2$, $W=1$, and $\phi = 0.2\phi_0$.}
\label{rev1}
\end{figure}

\subsubsection{Current with filling factor $N_{e}$ and system size $N$}

For three alternative circumstances between the hopping integrals, the nature of the current with filling factor ($N_e$) and system size ($N$) are shown in Fig.~\ref{f6} in two separate columns. Three distinct colored curves, red, green, and blue, indicate the dependency for $\nu=0$, $0.5$, and $0.8$, respectively. Now, if we concentrate on the correlation between the current and filling factor, we can see that for $t_1=t_2$, the current exhibits a predictable pattern with $N_{e}$. At half-filling, the current reaches its maximum and then starts to decline on each side for $\nu=0$. For $\nu=0.5$, the current is again quite higher at half-filling, but for other values of $\nu$ ($\nu=0.8$), the current is greater at other fillings and, as can be seen from the blue curve in Fig.~\ref{f6}, the current decreases as it approaches the half-filling. The current becomes minimal at half-filling than the other filled conditions when the relationship between the hopping integrals is changed to $t_{1}\ne t_{2}$. Such a situation and non-trivial behavior at half-filling are due to the presence of SSH patterns combined with the variations of $\nu$. Another important feature can be observed by inspecting the $I$-$N_e$ plots. For the disordered rings ($\nu=0.5$ and $\nu=0.8$), the current becomes vanishingly small when $N_e$ is either relatively low or high. This behavior arises solely due to the presence of nearly flat energy levels at both edges of the energy spectrum, as discussed earlier. Thus, by adjusting the filling factor, we can effectively control the current in the ring. 

We now proceed to the right column of Fig.~\ref{f6}, to illustrate the effect of system size on flux-driven circular current. Three different cases are considered depending on the hopping strengths $t_1$ and $t_2$, and in each diagram, three distinct values of $\nu$ are taken into account. For any situation, we find that the current decreases with the ring size, and the rate of fall becomes quite large for the dimerized rings compared to the ring with uniform hopping. The reduction of current with ring size is quite common, and has also been reported in other related works available in the literature. This is a direct consequence of the quantum interference effect. The key point is that we cannot expect a circular current for large enough ring sizes, as it is a mesoscopic phenomenon. 

\subsubsection{Current and eigenstate resolved NPR with $W$}

In Fig.~\ref{rev2}, we plot the persistent current at half-filling along with the eigenstate-index resolved NPR spectrum as a function of disorder strength for a 100-site ring. Although the persistent current primarily depends on the slope of the energy eigenvalues to flux, it can also be correlated with the NPR-resolved spectrum. The three rows, from top to bottom, correspond to three different conditions for the hopping integrals. For the case where $t_1 = t_2$, the persistent current decreases as $W$ increases. A similar trend is observed in the NPR spectrum, where the off-white region (indicating high NPR) is prominent at lower $W$, but as $W$ increases, this region shrinks while the dark reddish spectrum (low NPR) becomes more dominant. Since a high NPR signifies an extended phase associated with high conductivity (and thus high current), whereas a low NPR indicates an insulating phase with low conductivity, the decreasing persistent current with increasing $W$ is expected. For the other two cases where $t_1 \ne t_2$, the current exhibits significant fluctuations across a broad range near $W = 1.5$. This behavior can also be understood from the eigenstate-index resolved NPR spectrum. Initially, at low $W$, the NPRs are high, indicating extended states. 
However, as we approach half-filling, a gapped spectrum appears. When $W = 0$, the system corresponds to a dimerized lattice with two separate bands, a feature we have verified using the energy band diagram (though not shown here). Such a gapped spectrum leads to low conductivity 
and, consequently, a suppressed current. Around $W = 1.5$, if we examine the middle of the spectrum, we observe high-NPR regions mixed with low-NPR regions (white spots within a reddish background). This region represents an intermediate phase, where both extended and localized states coexist, leading to oscillatory behavior in the persistent current, fluctuating between high and low values around $W = 1.5$.

\subsubsection{Eigenvalues resolved NPR with $W$}

In our previous analysis, we examined the NPR resolved by the energy eigenstate index. To provide a more comprehensive perspective, in Fig.~\ref{revband} we present the NPRs as functions of energy eigenvalues and disorder strength $W$. The yellow regions correspond to high NPR values, indicative of extended states, while the dark blue regions represent lower NPR values, signifying localized states. When the hopping parameters are equal, i.e., $t_1 = t_2 = 1$, no sub-bands are present, and the extended region, marked by yellow, continuously diminishes with increasing $W$. This observation is consistent with the suppression of current as the disorder increases. However, for $t_1 \neq t_2$, the system initially exhibits two distinct sub-bands, as expected for a pristine SSH model at $W=0$. With increasing disorder, these sub-bands gradually merge, leading to an enhancement in conducting properties. This merging is accompanied by the appearance of a dark reddish hue blending into the off-white regions, particularly in areas where peaks and dips emerge in the current spectrum as a function of $W$. The presence of this mixed-color region signifies intermediate states that exhibit a coexistence of extended and localized phases as disorder strength increases, directly influencing the oscillatory nature of the current, which alternates between high and low values.

A comprehensive analysis of both Fig.~\ref{rev2} and Fig.~\ref{revband} reveals a strong correlation between the behavior of the persistent current and the NPR spectrum, whether resolved by eigenstate index or energy. This correlation underscores the crucial role of extended and localized states in governing transport properties. The observed trends highlight how disorder, hopping asymmetry, and flux collectively influence quantum transport in quasiperiodic systems. 

\vskip 0.25cm
\noindent
$\blacksquare$ {\bf Spin current:}
In Fig.~\ref{rev1}, we present the currents for up and down spin electrons along with their corresponding charge and spin currents in the non-interacting case. For the half-filled system (depicted in red), there are $8$ up-spin and $8$ down-spin electrons. At half-filling, the ground state energies of up and down spin electrons are identical, leading to equal currents for both spin components. As a result, the total charge current is simply the sum of the two spin currents, making it twice the individual spin currents. However, since the spin current is defined as the difference between up and down spin currents, it vanishes in this case.
In contrast, for the non-half-filled scenario, the up and down spin electrons occupy different energy states, resulting in unequal currents for the two spin components. This disparity arises due to the difference in their respective ground state energies. Consequently, both charge and spin currents acquire finite values, highlighting the role of the filling factor in determining spin transport properties in the system. 

\subsection{Interacting case}

Now, we start discussing the results for the interacting rings, where all the results are worked out based on the standard HF FM scheme.

\subsubsection{Variation of current with Hubbard strength $U$}

\begin{figure}[ht]
\noindent
{\centering\resizebox*{6.5cm}{9cm}{\includegraphics{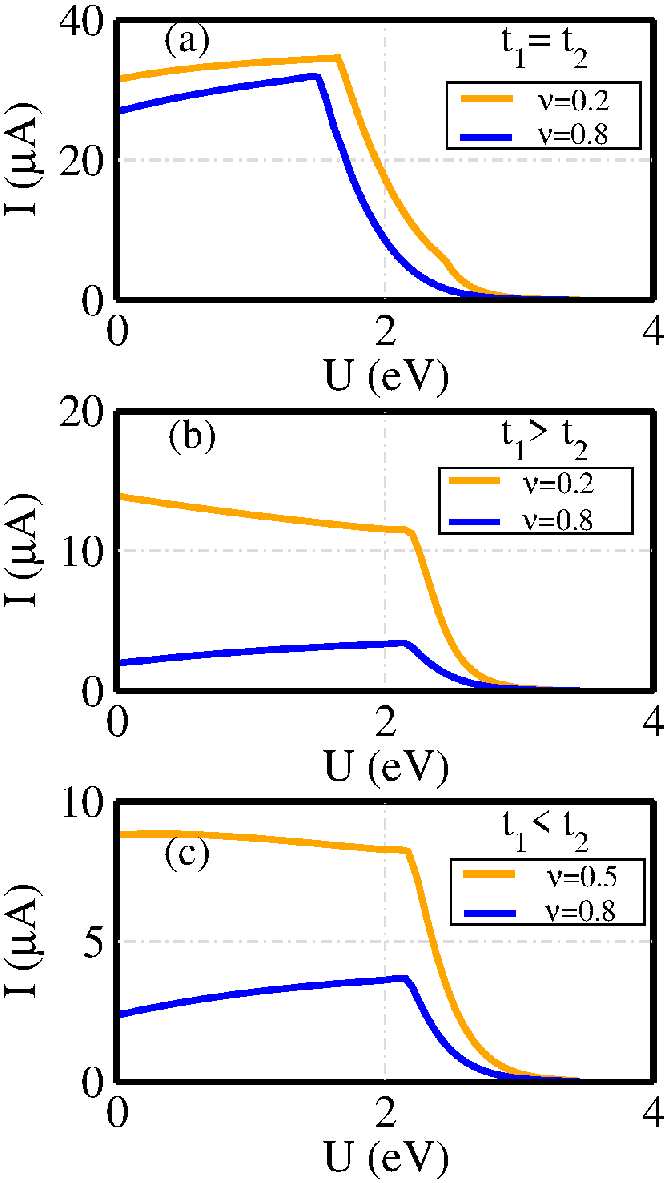}\par}}
\caption{(Color online). Dependence of current on the Coulomb interaction strength $U$ for a $16$-site ring, at half-filling. The results are computed for two distinct values of $\nu$, setting $W=1$.}
\label{f7}
\end{figure}

In Fig.~\ref{f7} we plot the typical current, estimated at constant flux, and
disorder strength, against the Hubbard interaction strength $U$ for three different conditions among the hopping integrals $t_1$ and $t_2$. In each sub-figure, the two colored curves are associated with two different values (low and high) of $\nu$. For $t_{1}=t_{2}$, the current initially rises with $U$, reaching a maximum, then decreases, and eventually drops to zero for sufficiently large $U$.
\begin{figure}[ht]
\noindent
{\centering\resizebox*{7cm}{9cm}{\includegraphics{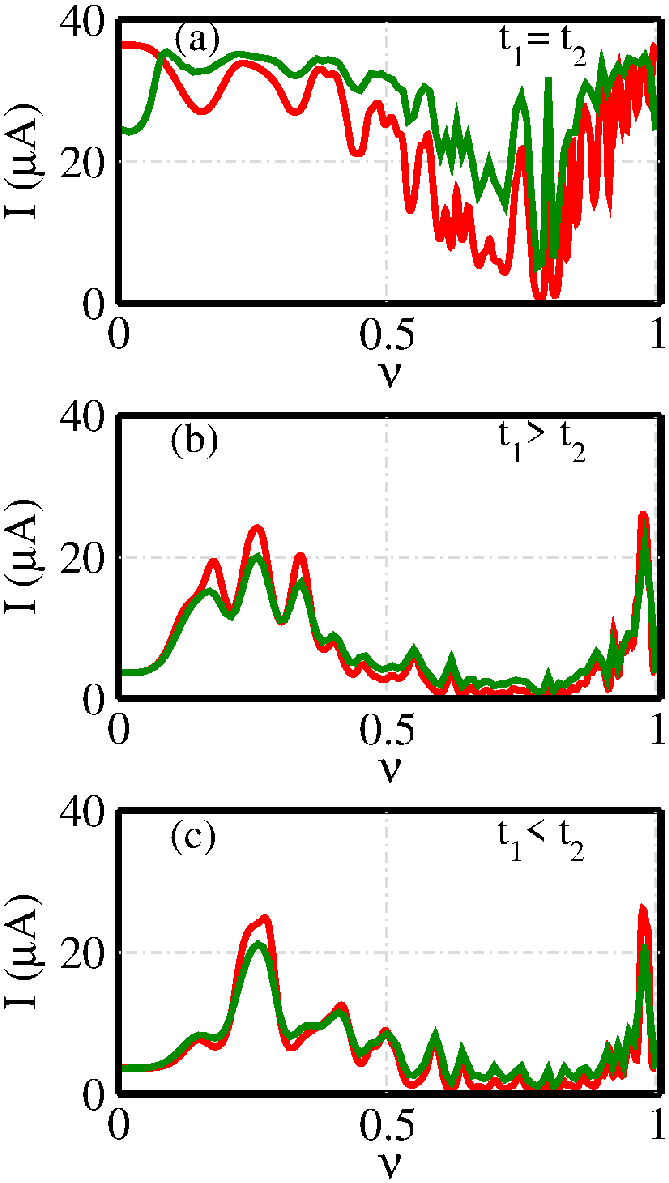}\par}}
\caption{(Color online). Circular current as a function of $\nu$ for a particular modulation strength ($W=1$), under three different conditions of $t_1$ and $t_2$. To directly compare the non-interacting and interacting cases, we use two values of $U$, $0$ and $1.5$, and the results are presented by the red and green curves, respectively. The ring size is fixed to $16$.}
\label{f8}
\end{figure}
\begin{figure*}[ht]
\noindent
{\centering\resizebox*{10cm}{9cm}{\includegraphics{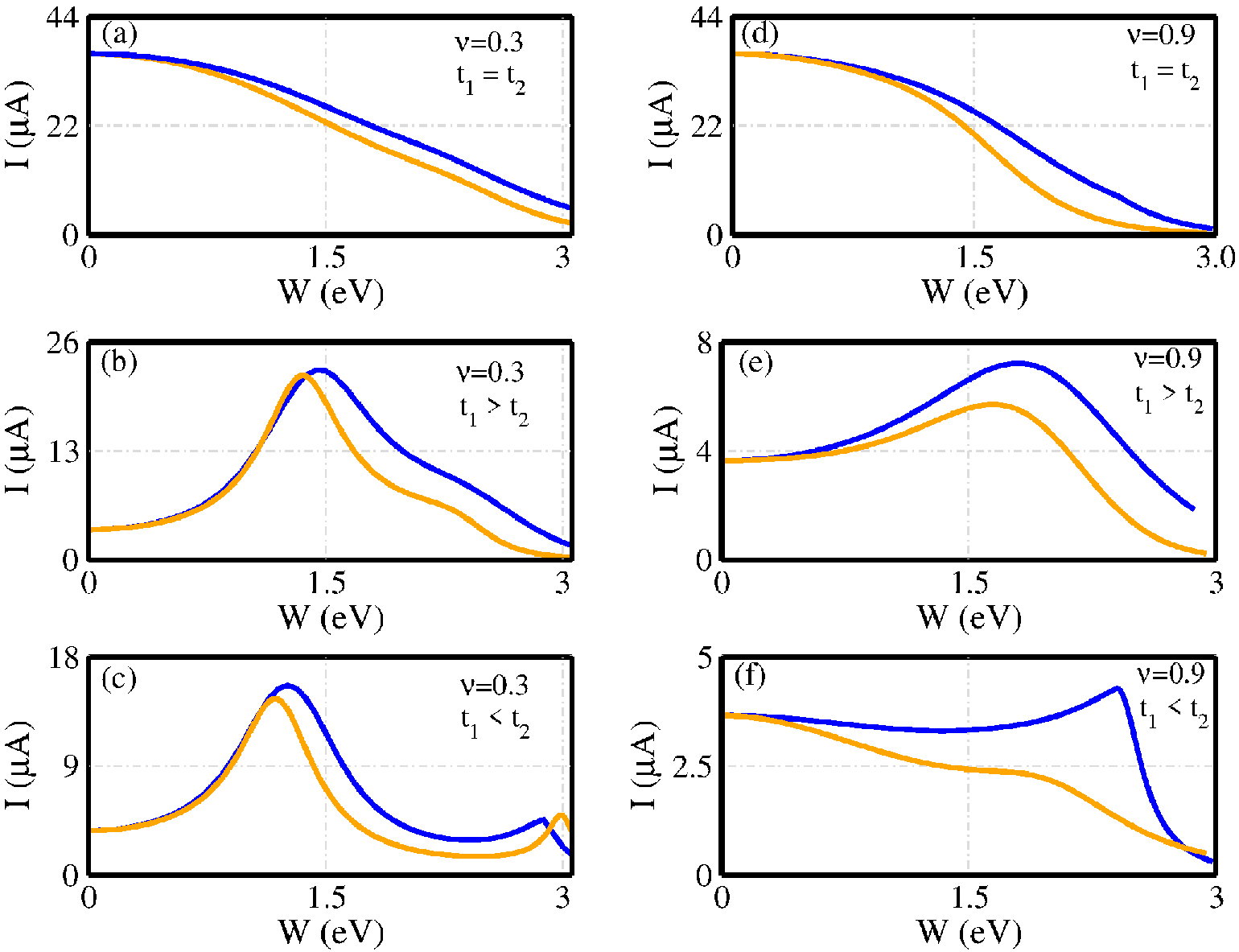}\par}}
\caption{(Color online). Variation of current with disorder strength $W$, for the half-filled band case, where the two columns are associated with two distinct values of $\nu$. The orange and blue curves correspond to $U=0$ and $0.5$, respectively. The result of the non-interacting ring is used to have a better comparison with the interacting one. Here, we set $N=16$.}
\label{f9}
\end{figure*}
The nature of the current is directly influenced by the interplay between disorder and electron-electron interactions. In a disordered ring, site energies vary, and when $U=0$, electrons with both spin orientations tend to localize at sites with lower potential, reducing their mobility through the ring. Introducing $U$ counteracts this localization, promoting electron movement and enhancing the current. As a result, the current initially increases with $U$. However, another crucial aspect must be considered. Since the system is half-filled, each site tends to accommodate a single electron, and this tendency strengthens with increasing $U$. As occupancy constraints grow, electron flow is hindered due to repulsive interactions, which prevent double occupancy and subsequently reduce the current. For sufficiently large $U$, nearly all sites become singly occupied, severely restricting the hopping of electrons with opposite spin, ultimately leading to a vanishing current. This scenario may arise for any value of $\nu$. However, when $t_{1} \neq t_{2}$, the pattern is more interesting and important as well. Regardless of the existence of quasi-periodic disorder, which is readily apparent from the plots, the current for both $t_{1}>t_{2}$ (or $t_{1}<t_{2}$) declines with $U$ from the beginning itself, for certain values of $\nu$ ($0.2$ for $t_{1}>t_{2}$ and $0.5$ for $t_{1}<t_{2}$). Hence, we can conclude that for those values of $\nu$, the influence of $U$ remains strong even in the lower range of values. However, for certain other values of $\nu$, the current is exhibiting a nearly identical pattern to what we see for $t_{1}=t_{2}$. We might anticipate a pattern similar to that in Fig.~\ref{f7}(a) in a typical quasi-periodic Hubbard system, but the system becomes even more interesting when we consider hopping dimerization. 
Thus, we can properly modulate the reliant nature of the current on Hubbard strength by adjusting the index $\nu$. 

\subsubsection{Dependency of current with $\nu$ in the presence of Coulomb interaction}

Figure~\ref{f8} shows how the current gets changed with $\nu$ in the presence of the Coulomb interaction. The variations for two distinct values of $U$ under three different situations between the hopping integrals are shown by the curves with red and green colors, respectively. As the system achieves intrinsic or perfect limits for those values, the current becomes maximum for $\nu=0$ and $1$. However, for other values of $\nu$, the configuration of the system causes the amplitude of the current to fluctuate with $\nu$. Therefore, we can have a situation where a higher or lower current can be obtained by selectively adjusting the parameter $\nu$. The current is now greater for certain other values of $\nu$ when the Hubbard interaction is taken into consideration instead of being at its maximum for $\nu=0$ or $1$. Another feature is that, for certain values of $\nu$, the current for non-zero $U$ predominates, while for $\nu=0$ (or $1$), the current for non-zero $U$ is smaller than the current for the interaction-free case. This occurs due to the interplay between the Coulomb interaction and quasi-periodicity. The hopping mechanism is enhanced more than the ideal circumstances ($\nu=0$ or $1$) as a result of such a combined effect. The current for some $\nu\ne 0$ or $1$ is greater than the values we get for $\nu=0$ $(1)$, as can be seen if we concentrate on the other two graphs for $t_{1}>t_{2}$ and $t_{1}<t_{2}$. In these situations, once we take into account the finite Hubbard strength, the current for $U=0$ is greater than $U=1.5$ for a certain range of $\nu$, while the converse is true for some other ranges of $\nu$. A justification for such a situation is given below. When dealing with dimerized situations, the hopping procedure gets suppressed and improved for various ranges of $\nu$. Thus, the hopping dimerization, quasi-periodicity, and Coulomb interaction all operate collectively to cause the current to increase and decrease, as can be seen from the plots given in Figs.~\ref{f8}(b) and (c). 
\begin{figure}[ht]
\noindent
{\centering\resizebox*{7cm}{7.5cm}{\includegraphics{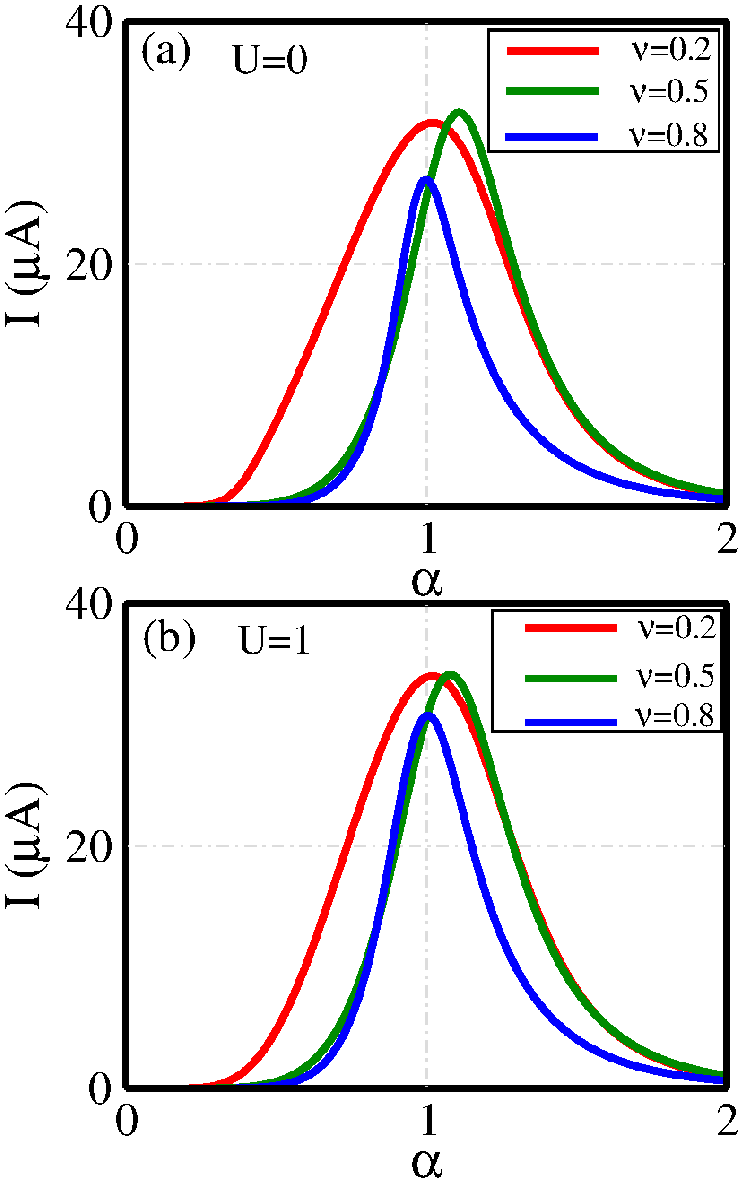}\par}}
\caption{(Color online). Current as a function of $\alpha$ ($=t_2/t_1$), in the limit of half-filling, for the interacting-free and
interacting rings. In each sub-figure, three different curves are given that are associated with three distinct values of $\nu$.
Here, we set $W=1$ and $N=16$.}
\label{f10}
\end{figure}

\subsubsection{Dependence of current on the modulation strength $W$}

The study of the dependence of current on $W$ in either the presence or absence of $U$ is always curious since disorder and Hubbard 
interaction constitute competition with one another. With the orange and blue curves, Fig.~\ref{f9} illustrates the variation of 
current with disorder strength for two distinct values of $U$ ($0$ and $0.5$). Here, we take two different values of $\nu$ 
($0.3$ and $0.9$) and divide the output into two columns. Although the present value of the current for $U=0.5$  is greater than the same for $U=0$, the current for $t_{1}=t_{2}$ invariably declines with $W$. The concurrent effect of $U$ and $W$ is what causes such an increase in current for finite $U$. When $t_{1}\ne t_{2}$ occurs, the situation changes. The pattern is essentially identical for both $\nu=0.3$ and $0.9$, but once $U$  becomes finite, the current is greater, as previously mentioned. Initially, the current rises and then falls for higher $W$. Another notable finding is the presence of two peaks when $t_{1}<t_{2}$ and $\nu=0.3$, with the second peak appearing around $W=3$. Interestingly, this second peak shifts to lower values of $W$ when $U$ becomes $0.5$. Furthermore, the current reduces with $W$ for $t_{1}<t_{2}$ at $\nu=0.9$ and $U=0$, but when considering a finite $U$, the current initially increases with $W$ even if it further gets lowered over a higher range of $W$. These graphs make it abundantly evident that the increase and decrease of current with $W$ in both the absence and presence of $U$ is the result of the combined effects 
of $t_{1}\ne t_{2}$, disorder (slowly varying), and Hubbard interaction.

\subsubsection{Variation of current with $\alpha$}

The dependence of current on the factor $\alpha$ (where $\alpha$ represents the ratio between $t_{2}$ and $t_{1}$) is noteworthy, as the SSH pattern present in our system contributes to several non-trivial properties. Three alternative values of $\nu$ are represented by the 
red, green, and blue curves, while in the plot (Fig.~\ref{f10}), two separate sub-plots indicate the presence and absence of the Hubbard interaction, respectively. The current is almost negligible when $t_{2}$ is significantly smaller than $t_{1}$ because the alternative bonds related to $t_2$ are ready for breakdown, and the hopping process is suppressed. The current increases to its maximum level when these connections are re-established, i.e., for the situation when $t_{2}$ and $t_{1}$ are comparable to each other. Although the occurrence of the peak might vary depending on $\nu$, it roughly takes place when the dimerized hopping strengths are quite comparable. Once again, when $t_{2}$ is much larger than $t_{1}$, the bonds connected to $t_{2}$ contract disrupt the hopping process at the bonds connected to $t_{1}$. As a result, the current is once again lowered to zero. Thus, we can say that the interplay among the Hubbard interaction, hopping dimerization, and the slowly varying potential play a significant impact on circular current and can be utilized further in other systems as well.  

\begin{figure}[ht]
\noindent
{\centering\resizebox*{8cm}{7cm}{\includegraphics{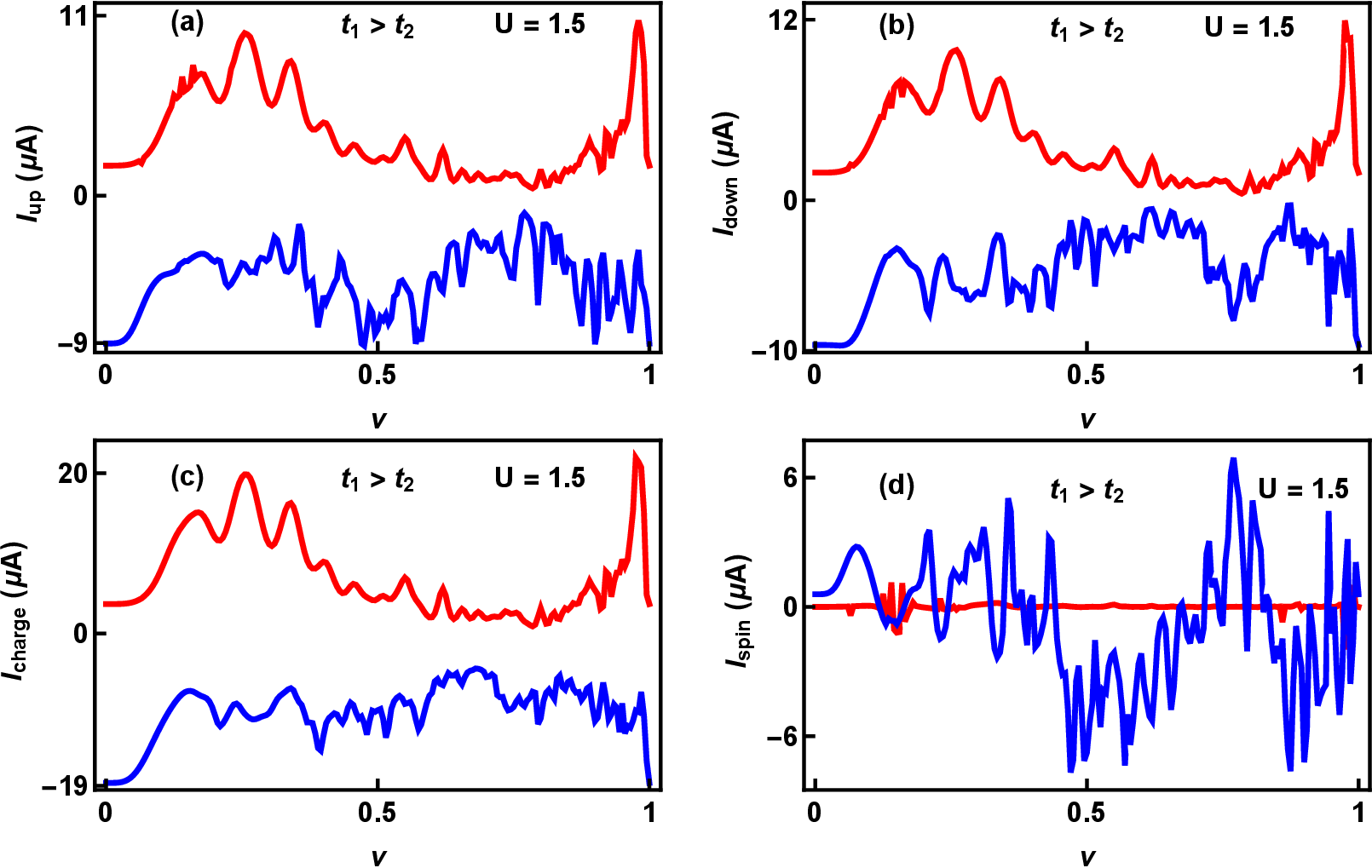}\par}}
\caption{(Color online). Up and down spin currents, along with their corresponding charge and spin currents as a function of $\nu$, for both half-filled (red) and non-half-filled (blue) cases, in the presence of electron-electron interaction ($U = 1.5$). The results are shown for a $16$-site ring with $W=1$ and $\phi = 0.2\phi_0$, considering the condition $t_1>t_2$. For the non-half-filled ring, we take $7$ up-spin and $5$ down-spin electrons, whereas for the half-filled ring equal number of up and down-spin electrons exist.}
\label{rev3}
\end{figure}

\vskip 0.25cm
\noindent
$\blacksquare$ {\bf Spin current:}
In Fig.~\ref{rev3}, we present the up and down spin currents along with the charge and spin currents
for a 16-site ring at both half-filling (represented in red) and non-half-filling (represented in blue), as indicated in the figure for $U=1.5$. For the half-filled case, the up and down spin currents are no longer equal, but they exhibit some variations when $U$ is switched on to $1.5$. For the non-half-filled case, it has already been demonstrated that the up and down spin currents differ even at $U=0$, and here, we 
observe that their magnitudes change further with the introduction of $U$. Focusing on the charge and spin currents, we notice that even at half-filling, the spin current is not zero over the entire range of $\nu$, except at $\nu=0$ and $1$. This contrasts with the $U=0$ case, where the spin current vanishes completely. In the half-filled case where an equal number of up and down spin electrons exist, spin current emerges under two conditions: (i) $U$ is non-zero and (ii) $W$ is non-zero. If either of these two is zero, the spin current vanishes. The underlying physics is very interesting, and it is explained as follows. In the half-filled limit, the system becomes `antiferromagnetic' (AFM) in the presence of electron-electron interaction. Because of this AFM nature, the effective up and down spin sub-Hamiltonians are symmetric to each other, resulting in identical energy spectra, and thus the current components. Since both the current components are identical, the spin current does not appear. The symmetry between the two sub-Hamiltonians can be broken once the disorder is included. For $\nu=0$ and $1$, the symmetry exists, and hence there is no spin current, whereas for all other values of $\nu$, a finite spin current appears. The combined effect of $\nu$ and electron-electron interaction on spin current is very significant and may have important implications for spintronic applications, where controlling spin currents is crucial for designing efficient spin-based devices. The observed behavior underscores the role of interaction-driven spin dynamics, which could be leveraged to optimize spin current generation and manipulation in correlated electronic systems. 

\subsection{Validation of MF Results – Comparison with Exact Diagonalization of full MB Hamiltonian}

To ensure the reliability of the MF results, it is crucial to verify their accuracy,
\begin{figure}[ht]
\noindent
{\centering\resizebox*{7cm}{9cm}{\includegraphics{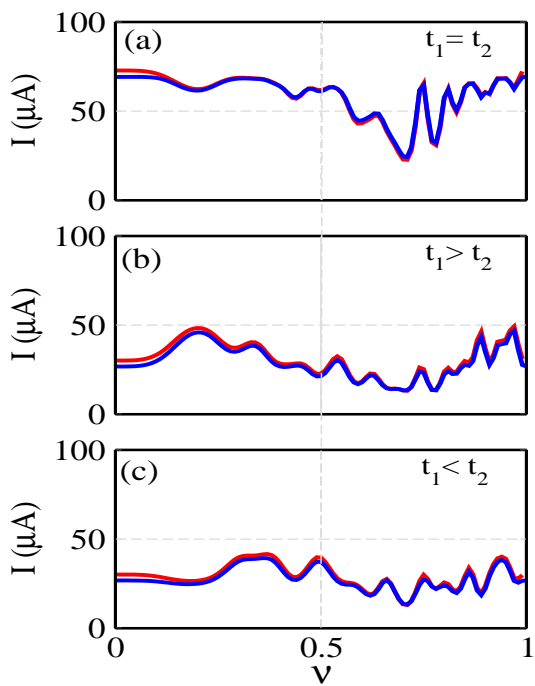}\par}}
\caption{(Color online). Comparison of mean-field and exact diagonalization results for an $8$-site ring at half-filling with $W = 1$ and $U = 1$. The red curve represents the mean-field calculations, while the blue curve corresponds to the exact diagonalization results.}
\label{rev4}
\end{figure}
as the MF method is based on certain approximations, and it may overlook certain essential things. For this purpose, Fig.~\ref{rev4} compares mean-field calculations and exact diagonalization (ED) of the full MB Hamiltonian, where we depict the magnitude of the typical current at half-filling as a function of $\nu$. Both MF and ED results are presented for different hopping scenarios, akin to those examined earlier. As highlighted before, a major limitation of the exact diagonalization method in many-body systems stems from the computational complexity of determining energy eigenvalues. This difficulty arises due to the exponential growth of the Hamiltonian matrix with increasing system size and the number of up and down spin electrons. Consequently, computational constraints restrict our analysis to smaller systems. Here, we consider a ring consisting of $8$ sites containing four spin-up and four spin-down electrons, representing a half-filled system. The comparison in Fig.~\ref{rev4} shows a strong agreement between the MF and ED results over a wide range of $\nu$, with both methods producing nearly identical curves. This consistency confirms the reliability of the MF approximation, reinforcing confidence in its predictions for the studied parameters. 

\vskip 0.2cm
\noindent
\subsection{Experimental realization}

The scientists conduct many experiments on the measurement of persistent current~\cite{r38,r47,r65} in various loop geometries. 
Additionally, the topological phase transition was investigated by experimental work with the SSH model~\cite{r32,r66}. Several apparent transport features in organic polymers that mimic the topological excitations in the SSH model are seen from such analysis~\cite{r32}. In our approach, the experimental manifestation may likewise be achieved by arranging the atomic sites so they display a slowly fluctuating pattern. Changes in the distance between the atomic sites may also be used to monitor two different types of hopping integrals. After designing a loop structure from such an atomic configuration, we can use the SQUID technique to measure the current through the computation of the magnetic dipole moment by immersing the system in a magnetic field. To create an SSH ring, varying hopping integral values can be attained by employing either two distinct atoms in successive sites or by arranging atoms with differing spacings in consecutive bonds. Manipulating the distance between two atoms with precise control can be accomplished using the atom manipulation technique. Quasi-periodicity can be achieved by utilizing two counter-propagating laser beams to trap atoms within those profiles.

\section{Closing Remarks}

The main goal of this work is to study various captivating aspects of flux-driven circular charge and spin currents in a quantum ring, including the inclusive impact of slowly varying site energies, hopping dimerization, and Coulomb interaction. We separately take into consideration non-interacting and interacting systems to precisely extract the interesting features. With two distinct hopping integrals, disorder 
strength, and different choices of slowly varying parameters, we monitor the current for non-interacting systems, while taking into consideration the combined influence of those parameters together with $U$, the investigation in the interacting system is done. Simulating the quantum ring within a tight-binding framework, we compute the energy eigenvalues by diagonalizing the Hamiltonian, and we find the current following the well-known derivative method. The effect of electron-electron interaction is incorporated through the Hartree-Fock mean-field scheme. All the results are carefully investigated under different input conditions. The key findings are as follows. \\
$\bullet$  Appearance of nearly flat energy levels around the edges of the energy band spectrum under suitable parameter values.\\
$\bullet$ For $t_{1}=t_{2}$, the current is consistently lower in disordered cases than the ordered ones. But, for $t_{1}\ne t_{2}$, the situation mostly reverses.\\
$\bullet$ By changing the index $\nu$, the current magnitude can be modified selectively. \\
$\bullet$ The re-entrant delocalization signature may be replicated by increasing the current with $W$. \\
$\bullet$ The role of the filling factor becomes more prominent for $t_{1}\ne t_{2}$ cases.\\
$\bullet$ When Coulomb interaction is present, increasing and decreasing behavior of current with disorder 
strength and a slowly varying index can be seen. The enhancement is especially apparent for a lesser Hubbard strength $U$. \\
$\bullet$ Up to a critical limit of $U$, the current increases or decreases monotonically depending on the choices of $t_{1}$ and $t_{2}$.\\  
$\bullet$ The inter-relationship among hopping integrals, Hubbard interaction, and slowly varying disorder interacts in a way that plays a key role in the growth and diminution of current, as is seen through the plot of current with $\alpha$. \\
$\bullet$ In the limit of half-filling, a circular spin current appears when both $U$ and $W$ are finite. For other filling cases, a spin current may emerge regardless of $U$ and $W$. \\
$\bullet$ From the accuracy check, it is confirmed that the MF results are quite comparable to those obtained from the ED of the full many-body Hamiltonian.

In the end, we can say that the present study brings several interesting and important features as a result of the interplay, especially between the slowly varying site potentials and the hopping dimerization, that can be utilized in different contexts like studying electron and spin-dependent transport phenomena, both in isolated and open quantum systems.

\section*{ACKNOWLEDGMENTS}

DL acknowledges partial financial support from the Centers of Excellence with BASAL/ANID financing, AFB220001, CEDENNA. We would like to thank all the reviewers for their constructive criticisms, valuable comments, and suggestions, which have greatly improved the quality of our work. 

\section*{DATA AVAILABILITY STATEMENT}

The data that support the findings of this study are available upon reasonable request from the authors.

\section*{DECLARATION}

{\bf Conflict of interest} The authors declare no conflict of interest.

\end{document}